\documentclass[prd,twocolumn,showpacs,floatfix,amsmath,nofootinbib,amssymb,floatfix]{revtex4}
\usepackage{graphicx,color,dcolumn,booktabs,bm,multirow}
\usepackage{longtable,lscape}
\usepackage{txfonts}
 \usepackage{enumitem}
\usepackage{overpic}
\usepackage{amssymb}
\usepackage{indentfirst}
\usepackage{feynmf}   
\usepackage{slashed}  
\usepackage{cases}
\usepackage{color}
\usepackage{multirow}
\usepackage{epstopdf}
\usepackage{graphicx,color,dcolumn,booktabs,bm}

\usepackage[colorlinks,
            citecolor=blue,
            anchorcolor=red,
            menucolor=red,
            linkcolor=red,
            filecolor=red,
            runcolor=red,
            urlcolor=blue,
            frenchlinks=red]{hyperref}
\usepackage{amsmath}
\usepackage{epsfig}
\usepackage{graphicx}
\usepackage{txfonts}
\def\r{\boldsymbol{r}}
\def\R{\boldsymbol{R}}

\begin{document}

\title{Investigation of the bottom analog of the $Z_{cs}(3985)$ state}

\author{Xuejie Liu$^1$}\email[E-mail: ]{1830592517@qq.com}
\author{Dianyong Chen$^{1,6}$\footnote{Corresponding author}}\email[E-mail:]{chendy@seu.edu.cn}
\author{Hongxia Huang$^2$}\email[E-mail:]{hxhuang@njnu.edu.cn}
\author{Jialun Ping$^2$}\email[E-mail: ]{jlping@njnu.edu.cn}
\author{Xiaoyun Chen$^{3}$}\email[E-mail:]{xychen@jit.edu.cn}
\author{Youchang Yang$^{4,5}$}\email[E-mail:]{yangyc@gues.edu.cn}
\affiliation{$^1$School of Physics, Southeast University, Nanjing 210094, P. R. China}
\affiliation{$^2$Department of Physics, Nanjing Normal University, Nanjing 210023, P. R. China}
\affiliation{$^3$College of Science, Jinling Institute of Technology, Nanjing 211169, P. R. China}
\affiliation{$^4$School of Science, Guizhou University of Engineering Science, Bijie 551700, P. R. China}
\affiliation{$^5$School of Physics and Electronic Science, Zunyi Normal University, Zunyi 563006, P. R. China}
\affiliation{$^6$Lanzhou Center for Theoretical Physics, Lanzhou University, Lanzhou 730000, P. R. China}

\begin{abstract}
Motivated by the recent discovery of the hidden charm exotic state with strangeness by the BESIII and LHCb Collaborations, we study the $S$ wave strange hidden bottom tetraquark in two kinds of quark models. Both meson-meson and diquark-antidiquark configurations are taken into account. The numerical results indicate that there is no bound state in both quark models. However, several resonance states have been predicted. Three resonance states with $I(J^{P})=\frac{1}{2}(0^{+})$ are found, the energy ranges of which are $10479\sim 10550$, $10528\sim 10632$, and $10597\sim 10681$ MeV, respectively. Three resonance states with $I(J^{P})=\frac{1}{2}(1^{+})$ are predicted to be located in $10491\sim 10675$, $10502\sim 10679$, and $10522\sim 10723$ MeV, respectively. Moreover, there also exist a resonance with $I(J^{P})=\frac{1}{2}(2^{+})$ and the mass is estimated to be $10531\sim 10680$ MeV. All these predicted states in the present work should be accessible for the further experiments in LHCb
\end{abstract}

\date{\today}

\pacs{13.75.Cs, 12.39.Pn, 12.39.Jh}
\maketitle

\setcounter{totalnumber}{5}

\section{\label{sec:introduction}Introduction}

In the last two decades, many experimental efforts have been made to search for QCD exotic states and great progresses have been archived (for recent review, we refer to Ref~\cite{Chen:2016qju,Hosaka:2016pey,Lebed:2016hpi,Esposito:2016noz,Guo:2017jvc,Ali:2017jda,Olsen:2017bmm,Karliner:2017qhf,Yuan:2018inv}). Among the newly observed hadron states, a series of charged heavy-quarkonium-like states are particularly interesting since they are definite good candidates of multiquark states, which have stimulate great interests of both theorists and experimentalists.

In the bottom sector, the Belle Collaboration reported two bottomonium-like states, $Z_b(10610)$ and $Z_b(10650)$, in the invariant mass of $\pi^\pm \Upsilon(nS), \ (n=1,2,3)$ and $\pi^\pm h_b(mP),\ (m=1,2)$ of the dipion decays of $\Upsilon(5S)$ in the year of 2011~\cite{Belle:2011aa}. And later, the Belle Collaboration also observed these two bottomonium-like states in the open bottom decays of $\Upsilon(5S)$~\cite{Belle:2012koo,Belle:2015upu}. From the observed channels, one can find the most possible quark components of $Z_b$ states are $b\bar{b} q\bar{q}$ with $q=(u,d)$, thus these two bottomonium-like states can be considered as tetraquark states~\cite{Ali:2011ug,Esposito:2014rxa,Maiani:2017kyi}. Moreover, the observed masses of $Z_b(10610)$ and $Z_b(10650)$ are very close to the thresholds of $B^\ast \bar{B}$ and $B^\ast \bar{B}^\ast$, respectively, which indicate that $Z_b(10610)$ and $Z_b(10650)$ can be good candidates of deuteron-like molecular states composed of $B^\ast \bar{B}+c.c$ and $B^\ast \bar{B}^\ast$, respectively~\cite{Bondar:2011ev,Cleven:2011gp,Nieves:2011zz,Zhang:2011jja,Yang:2011rp,Ohkoda:2011vj,Li:2012wf,Ke:2012gm,Dias:2014pva}. Besides these exotic interpretations, some particular kinematical mechanisms, such as cusp effect~\cite{Swanson:2014tra} and initial single pion emission mechanism~\cite{Chen:2011pv,Chen:2012yr}, have also been proposed.

Considering the heavy quark symmetry, the charm analogs of $Z_b$ states were predicted theoretically in different scenarios~\cite{Liu:2009qhy,Aceti:2014kja,Zhang:2013aoa,Chakrabarti:2014dna,Navarra:2001ju,Aceti:2014uea, Maiani:2007wz,Chen:2010ze,Zhao:2014qva,Voloshin:2013dpa,Patel:2014vua,Faccini:2013lda,Deng:2015lca,Szczepaniak:2015eza,Chen:2011xk,Chen:2013coa,Swanson:2015bsa,Swanson:2014tra}. On the experimental side, In the year of 2013 the BESIII and Belle Collaborations reported a new structure named $Z_c(3900)$ in the invariant mass spectrum  $\pi^{\pm}J/\psi$ of the process $e^{+}e^{-}\rightarrow\pi^{+}\pi^{-}J/\psi$ at a center of mass energy of 4.260 GeV~\cite{BESIII:2013ris, Belle:2013yex}. Later on, the CLEO-c Collaboration also confirmed the existence of this charmonium-like state in the same process but at $\sqrt{s}=4170$ MeV~\cite{Xiao:2013iha}. By analyzing the helicity angle distributions of $e^+e^- \to \pi^+ \pi^- J/\psi$ process, the spin-parity of $Z_{c}(3900)$ was determined to be $J^{P}=1^{+}$~\cite{BESIII:2017bua}. In the same year, another charmonium-like state, named $Z_c(4020)$, was also reported by the BESIII Collaboration in the $h_c \pi^\pm$ invariant mass distributions of the process $e^+e^- \to \pi^+ \pi^- h_c$~\cite{BESIII:2013ouc}. Besides the hidden charm process, two charged charmonium-like states, $Z_c(3885)$  and $Z_c(4025)$, were observed in the $D^\ast \bar{D}$ invariant mass spectrum of the process $e^+ e^- \to (D^\ast \bar{D})^\mp \pi^{\pm}$~\cite{BESIII:2013qmu} and $D^\ast \bar{D}^\ast$ invariant mass spectrum of the process $e^+ e^- \to (D^\ast \bar{D}^\ast)^\mp \pi^{\pm}$~\cite{BESIII:2013mhi}, respectively. Similar to the cases of $Z_b(10610)$ and $Z_b(10650)$, these charmonium-like states are charged, and moreover, the observed masses of $Z_c(3900)/Z_c(3985)$ are near the threshold of $D^\ast \bar{D}$, while the masses of $Z_c(4020)/Z_c(4025)$ are close to the threshold of $D^\ast \bar{D}^\ast$. Thus, similar tetraquark~\cite{Maiani:2007wz,Chen:2010ze,Zhao:2014qva,Voloshin:2013dpa,Patel:2014vua,Faccini:2013lda,Deng:2015lca}, molecular~\cite{Liu:2009qhy,Aceti:2014kja,Zhang:2013aoa,Chakrabarti:2014dna,Navarra:2001ju,Aceti:2014uea} and special kinematical mechanism~\cite{Swanson:2014tra,Szczepaniak:2015eza,Chen:2011xk,Chen:2013coa,Swanson:2015bsa} explanations have been applied to interpret these $Z_c$ states.

\renewcommand\arraystretch{1.25}
\begin{table*}[htb]
\caption{The resonance parameters of $Z_b$, $Z_c$ and $Z_{cs}$ states and the corresponding thresholds. \label{Tab:ResPara}}
\begin{tabular}{cccc|cccc}
\toprule[1pt]
\multicolumn{4}{c|}{Charm Sector} &\multicolumn{4}{c}{Bottom Sector} \\
\midrule[1pt]
States& Mass (MeV) & Width (MeV)& Threshold &States& Mass (MeV) & Width (MeV)& Threshold \\
\midrule[1pt]
$Z_c(3900)$  &$3899\pm3.6\pm4.9$  &$46\pm10\pm20$  & \multirow{2}{*}{$D^\ast \bar{D}$}    &\multirow{2}{*}{$Z_b(10610)$} &\multirow{2}{*}{$10607.2\pm2.0$} &\multirow{2}{*}{$18.4\pm2.4$} &\multirow{2}{*}{$B^\ast \bar{B}$}\\
$Z_c(3885)$  &$3883.9\pm1.5\pm4.2$  &$24.8\pm3.3\pm11.0$  &     &\\
$Z_c(4020)$  &$4022.9\pm0.8\pm2.7$  &$7.9\pm2.7\pm2.6$  & \multirow{2}{*}{$D^\ast \bar{D}^\ast$}    &\multirow{2}{*}{$Z_b(10650)$} &\multirow{2}{*}{$10652.2\pm1.5$} &\multirow{2}{*}{$11.5\pm2.2$} &\multirow{2}{*}{$B^\ast \bar{B}^\ast$}\\
$Z_c(4025)$  &$4026.3\pm2.6\pm3.7$  &$24.8\pm5.6\pm7.7$  &     &\\
\midrule[1pt]
$Z_{cs}(3985)$  &$3982.5^{+1.8}_{-2.6}\pm2.1$  &$12.8^{+5.3}_{-4.4}\pm3.0$  & \multirow{2}{*}{$D_s^\ast \bar{D}$}    & \multirow{2}{*}{$Z_{bs}$} &\multirow{2}{*}{$?$} & &\multirow{2}{*}{$B^\ast \bar{B}^{(\ast)}$}\\
$Z_{cs}(4000)$  &$4003\pm6^{+4}_{-14}$  &$131\pm15\pm26$  &     &\\
\bottomrule[1pt]
\end{tabular}
\end{table*}

It is interesting to notice that the productions of the above charged bottomonium- and charonium-like states are accompanied by a pion, which is a chiral particle. By replace the pion with another chiral particle, such as kaon, one can construct the strange partner of $Z_c$ states, named $Z_{cs}$. Such kind of charmonium-like states with strangeness have been investigated in various model. For example, the systematically estimations in the relativistic quark model in Ref.~\cite{Ebert:2005nc} evaluate the masses of heavy tetraquarks with hidden charm and bottom, and the lowest tetraquark states with $J^P=1^{+}$ were predicted to be around 4 GeV. Moreover, By extending the initial single pion emission mechanism to the single chiral particle emission mechanism, the authors in  Ref.~\cite{Chen:2013wca} predicted the structure near the thresholds of $D_s^\ast \bar{D}$ and $D_s^\ast \bar{D}^\ast$ in the hidden charm dikaon decays of higher charmonia.

The observations of $Z_c(3900)$ and the theoretical predictions of its strange partner simulated the experimentalists to search for the charmonium-like states with strangeness in various processes. Recently, the BESIII Collaboration observed a new structure, named $Z_{cs}(3985)$, in the $K^{+}$ recoil-mass spectrum of the process $e^{+}e^{-}\rightarrow K^{+}(D_{s}^{-}D^{*0}+D^{*-}_{s}D^{0})$ at the center of mass energy $\sqrt{s}=4.681$ GeV~\cite{BESIII:2020qkh}. More recently, the LHCb Collaboration reported two charmonium-like states with strangeness , named $Z_{cs}(4000)^+$ and $Z_{cs}(4220)^+$, in the $J/\psi K$ invariant mass spectrum~\cite{LHCb:2021uow}, where the mass of the former one is similar to the one of $Z_{cs}(3985)$ observed by BESIII Collaboration. Similar to the case of $Z_b(10610)$ and $Z_c(3900)/Z_c(3885)$, the observed masses of $Z_{cs}(3985)/Z_{cs}(4000)$ are close to the threshold of $D_s^\ast \bar{D}$, the properties of $Z_{cs}(3985)/Z_{cs}(4000)$ have been investigated in molecular ~\cite{Meng:2020ihj,Yang:2020nrt,Sun:2020hjw,Wang:2020htx,Dong:2020hxe,Xu:2020evn,Yan:2021tcp,Wu:2021ezz,Ozdem:2021yvo,Liu:2020nge,Chen:2020yvq} and tetraquark ~\cite{Wan:2020oxt,Wang:2020iqt,Jin:2020yjn,Giron:2021sla,Chen:2021uou,Yang:2021zhe} scenarios.

In Table \ref{Tab:ResPara}, we collect the resonance parameters of the above discussed $Z_b$, $Z_c$ and $Z_{cs}$ states, where one can clearly find that all these heavy-quarkonium like states are close to the thresholds of a pair of $S$ wave heavy-light mesons. The charmonium-like states $Z_c(3900)/Z_{cs}(3985)$ and $Z_c(4020)/Z_c(4025)$, in a sense, can be considered as the charm analogs of $Z_b(10610)$ and $Z_b(10650)$, while $Z_{cs}(3985)/Z_{cs}(4000)$ can be the strange partner of $Z_c(3900)/Z_c(3985)$. From the table and considering the heavy quark symmetry, it naturally to ask the question whether there are the bottom analogs of $Z_c(3900)/Z_c(3985)$ or the strange partners of $Z_b(10610)$ and $Z_b(10650)$ near the thresholds of $B_s^\ast \bar{B}^{(\ast)}$. This problem has been investigated by many theorists through different methods, which can be categorized into tetraquark and molecular scenarios. In Ref~\cite{Meng:2020ihj}, with the heavy quark flavor symmetry, the states $Z_{bs}$ and $Z_{bs}^{'}$ were predicted with a mass around 10700 and 10750 MeV, respectively. By using chiral effective field theory, the possibility of $Z_{bs}$ states was analyzed in the $B^{*}_{s}\bar{B}/B_{s}\bar{B}^{*}$ and $B^{*}_{s}\bar{B}^{*}$ systems~\cite{Wang:2020htx}, and two $Z_{bs}$ states, the $Z_{bs}(10700)$ and $Z_{bs}(10745)$, with the $I(J^{P})=\frac{1}{2}(1^{+})$ were predicted. The coupled channels estimation in a chiral constituent quark model found two virtual states at 10691 MeV and 10739 MeV in the bottom-strange sector~\cite{Ortega:2021enc}, , which could be considered as the strange partners of the well established $Z_{b}(10610)^{\pm}$ and the $Z_{b}(10650)^{\pm}$ states. Moreover, the estimations in Refs~\cite{Azizi:2020zyq,Sungu:2020zvk,Cao:2020cfx} indicated that there exist hidden bottom and open strange tetraquark states.

In Ref.~\cite{Jin:2020yjn}, we investigate the $Z_{cs}$ in the chiral quark model (ChQM) and we find a resonance state with $J^{P}=0^{+}$, the energy of which is close to the one of $Z_{cs}(3985)$. Besides, two more new resonances were predicated in the vicinity of 4410 MeV, with $J^{P}=0^{+}$ and $J^{P}=1^{+}$, respectively, which could be tested by further experiments. In the present work, we extend our previous estimations to the bottom sector and make a thorough analysis of the hidden bottom tetraquark with strangeness in the framework of the chiral quark model (ChQM) and the quark delocalization color screening model (QDCSM) to check whether there are bottom analogs of the $Z_{cs}$.

This work is organized as follows. After introduction, we present the details of the theoretical frame work adopted in the present work, including the phenomenological models and the wave functions of the tetraqaurk states as well. The numerical results and  the relevant discussions are presented in Section~\ref{results} and the last section is devoted to  a short summary.

 \section{THEORETICAL FRAMEWORK}{\label{model}}

To evaluate the spectra of strange hidden bottom states, two different kinds of quark models are adopted, which are the chiral quark model and the quark delocalization color screening model, respectively. In the following, we present a review of these two models.

\subsection{Chiral quark  model}
The ChQM is constructed based on the fact of nearly massless current light quark. However, the nonzero light-quark mass leads to a spontaneous chiral symmetry breaking in QCD, and then the current quarks become dressed constituent quarks. In the chiral quark model, the Hamiltonian for a tetraquark system is
\begin{eqnarray}
H  =  \sum_{i=1}^4\left(m_i+\frac{p_i^2}{2m_i}\right)-T_{CM} +\sum_{j>i=1}^4
\left(V^{C}_{ij}+V^{G}_{ij}+V^{\chi}_{ij}+V^{\sigma_{a}}_{ij}\right),
\end{eqnarray}
where $T_{CM}$ is the kinetic energy of the center of mass, and the potential between quarks/antiquarks includes the Goldstone-boson exchange potentials, the perturbative one-gluon interaction, and a linear-screened confining potential. The concrete forms of these potentials are ~\cite{Valcarce:2005em},
\begin{widetext}
\begin{eqnarray}
V^{C}_{ij} & = & -a_{c} \boldsymbol{\lambda}^c_{i}\cdot \boldsymbol{
\lambda}^c_{j} ({r^2_{ij}}+V_{0_{ij}}), \label{sala-vc} \nonumber\\
V^{G}_{ij} & = & \frac{1}{4}\alpha^{ij}_{s} \boldsymbol{\lambda}^{c}_i \cdot
\boldsymbol{\lambda}^{c}_j
\left[\frac{1}{r_{ij}}-\frac{\pi}{2}\delta(\boldsymbol{r}_{ij})(\frac{1}{m^2_i}+\frac{1}{m^2_j}
+\frac{4\boldsymbol{\sigma}_i\cdot\boldsymbol{\sigma}_j}{3m_im_j})-\frac{3}{4m_im_jr^3_{ij}}
S_{ij}\right], \label{sala-vG} \nonumber \\
V^{\chi}_{ij} & = & V_{\pi}( \boldsymbol{r}_{ij})\sum_{a=1}^3\lambda
_{i}^{a}\cdot \lambda
_{j}^{a}+V_{K}(\boldsymbol{r}_{ij})\sum_{a=4}^7\lambda
_{i}^{a}\cdot \lambda _{j}^{a}
+V_{\eta}(\boldsymbol{r}_{ij})\left[\left(\lambda _{i}^{8}\cdot
\lambda _{j}^{8}\right)\cos\theta_P-(\lambda _{i}^{0}\cdot
\lambda_{j}^{0}) \sin\theta_P\right], \label{sala-Vchi1} \nonumber \\
V_{\chi}(\boldsymbol{r}_{ij}) & = & {\frac{g_{ch}^{2}}{{4\pi
}}}{\frac{m_{\chi}^{2}}{{\
12m_{i}m_{j}}}}{\frac{\Lambda _{\chi}^{2}}{{\Lambda _{\chi}^{2}-m_{\chi}^{2}}}}%
m_{\chi} \left\{(\boldsymbol{\sigma}_{i}\cdot
\boldsymbol{\sigma}_{j})
\left[ Y(m_{\chi}\,r_{ij})-{\frac{\Lambda_{\chi}^{3}}{m_{\chi}^{3}}}%
Y(\Lambda _{\chi}\,r_{ij})\right] \right.\nonumber \\
&& \left. +\left[H(m_{\chi}
r_{ij})-\frac{\Lambda_{\chi}^3}{m_{\chi}^3}
H(\Lambda_{\chi} r_{ij})\right] S_{ij} \right\}, ~~~~~~\chi=(\pi, K, \eta). \nonumber\\
V^{\sigma_{a}}_{ij} & = & V_{a_{0}}(
\boldsymbol{r}_{ij})\sum_{a=1}^3\lambda _{i}^{a}\cdot \lambda_{j}^{a}+V_{\kappa}(\boldsymbol{r}_{ij})
\sum_{a=4}^7\lambda_{i}^{a}\cdot \lambda _{j}^{a}+V_{f_{0}}(\boldsymbol{r}_{ij})\lambda _{i}^{8}\cdot \lambda_{j}^{8}+V_{\sigma}(\boldsymbol{r}_{ij})
\lambda _{i}^{0}\cdot \lambda _{j}^{0}, \label{sala-su3} \nonumber \\
V_{k}(\boldsymbol{r}_{ij}) & = & -{\frac{g_{ch}^{2}}{{4\pi }}}
{\frac{\Lambda _{k}^{2}}{{\Lambda _{k}^{2}-m_{k}^{2}}}}%
m_{k}\left[ Y(m_{k}\,r_{ij})-{\frac{\Lambda _{k}}{m_{k}}}%
Y(\Lambda _{k}\,r_{ij})\right] , ~~~~~~k=(a_{0}, \kappa, f_{0}, \sigma). \nonumber\\
S_{ij}&=&\left\{ 3\frac{(\boldsymbol{\sigma}_i
\cdot\boldsymbol{r}_{ij}) (\boldsymbol{\sigma}_j\cdot
\boldsymbol{r}_{ij})}{r_{ij}^2}-\boldsymbol{\sigma}_i \cdot
\boldsymbol{\sigma}_j\right\},\nonumber \\
H(x)&=&(1+3/x+3/x^{2})Y(x),~~~~~~
 Y(x) =e^{-x}/x. \label{sala-vchi2}
\end{eqnarray}
\end{widetext}
Where $\alpha_s$ is the strong coupling constant.
The coupling constant $g_{ch}$ for the chiral field is determined from the $NN\pi$ coupling constant through
\begin{equation}
\frac{g_{ch}^{2}}{4\pi }=\left( \frac{3}{5}\right) ^{2}{\frac{g_{\pi NN}^{2}%
}{{4\pi }}}{\frac{m_{u,d}^{2}}{m_{N}^{2}}}\label{gch}.
\end{equation}
More details of the chiral quark model can be found in Ref.~\cite{Valcarce:2005em}.

\subsection{Quark delocalization color screening model}
The quark delocalization color screening model (QDCSM) is an extension of the native quark cluster model~\cite{DeRujula:1975qlm, Isgur:1978xj, Isgur:1978wd, Isgur:1979be} and is also developed with aim of addressing multiquark systems. The detail of QDCSM can be found in Refs.~\cite{Wang:1992wi, Wu:1996fm, Huang:2011kf}.  The Hamiltonian of QDCSM is almost the same as the one of ChQM but with two modifications~\cite{Wu:1996fm, Huang:2011kf, Ping:1998si, Wu:1998wu, Pang:2001xx, Huang:2015uda}. Firstly, there is no $\sigma$-meson exchange in QDCSM, and secondly, the screened color confinement is used between quark pairs resident in different clusters. That is
\begin{equation}
V_{ij}^{C}=\left \{ \begin{array}{ll}
-a_{c}\boldsymbol{\mathbf{\lambda}}^c_{i}\cdot
\boldsymbol{\mathbf{ \lambda}}^c_{j}~(r_{ij}^2+ V_{0_{ij}}) &
  \mbox{if \textit{i}th and \textit{j}th quark} \\
  &  \mbox{in the same cluster,} \\
-a_{c}\boldsymbol{\mathbf{\lambda}}^c_{i}\cdot
\boldsymbol{\mathbf{
\lambda}}^c_{j}~(\frac{1-e^{-\mu_{ij}\mathbf{r}_{ij}^2}}{\mu_{ij}}+
V_{0_{ij}}) & \mbox{otherwise,} \end{array} \right.\label{QDCSM-vc}
\end{equation}
where the color screening constants $\mu_{ij}$ are determined by fitting the deuteron properties, $NN$ scattering
phase shifts and $N\Lambda$, $N\Sigma$ scattering cross-sections~\cite{Chen:2011zzb,Ping:1993me,Wang:1998nk}. The concrete values of $\mu_{ij}$ are fitted to be $\mu_{qq}=0.45$, $\mu_{qs}=0.19$ and
$\mu_{ss}=0.08$, which satisfy the relation, $\mu_{us}^{2}=\mu_{uu}\mu_{ss}$. When extending to the heavy-quark case, the estimations in Ref~\cite{Huang:2015uda} indicated that the dependence of the parameter $\mu_{cc}$ is not very significant in the calculations of the $P_{c}$ states by taking $\mu_{cc}$ from 0.0001 to 0.01. Due to the heavy quark flavor symmetry, the behavior of the bottom quark should be similar to the one  of the charm quark. Herein, we take $\mu_{bb}=0.01$. Then $\mu_{bu}$ and $\mu_{bs}$ are obtained by the relation $\mu^{2}_{bu}=\mu_{bb}\mu_{uu}$ and $\mu^{2}_{bs}=\mu_{bb}\mu_{ss}$, respectively. Besides, according to Ref~\cite{Huang:2011kf}, the phenomenological color screening confinement is effective description of the hidden color channel coupling, so the hidden color channel of the tetraquark system in QDCSM cannot be included.

The single-particle orbital wave functions in the ordinary quark cluster model are the left and right centered single Gaussian functions, which are,
\begin{eqnarray}\label{wave2}
\phi_\alpha(\boldsymbol {S_{i}})=\left(\frac{1}{\pi
b^2}\right)^{\frac{3}{4}}e^ {-\frac{(\boldsymbol {r_{\alpha}}-\frac{1}{2}\boldsymbol
{S_i})^2}{2b^2}},
 \nonumber\\
\phi_\beta(-\boldsymbol {S_{i}})=\left(\frac{1}{\pi
b^2}\right)^{\frac{3}{4}}e^ {-\frac{(\boldsymbol {r_{\beta}}+\frac{1}{2}\boldsymbol
{S_i})^2}{2b^2}} .
 \
\end{eqnarray}
The quark delocalization in QDCSM is realized by writing the single-particle orbital wave function as a
linear combination of the left and right Gaussians, which are,
\begin{eqnarray}
{\psi}_{\alpha}(\boldsymbol {S_{i}},\epsilon) &=&
\left({\phi}_{\alpha}(\boldsymbol{S_{i}})
+\epsilon{\phi}_{\alpha}(-\boldsymbol{S_{i}})\right)/N(\epsilon),
\nonumber \\
{\psi}_{\beta}(-\boldsymbol {S_{i}},\epsilon) &=&
\left({\phi}_{\beta}(-\boldsymbol{S_{i}})
+\epsilon{\phi}_{\beta}(\boldsymbol{S_{i}})\right)/N(\epsilon),
\nonumber \\
N(\epsilon)&=&\sqrt{1+\epsilon^2+2\epsilon e^{{-S}_i^2/4b^2}}.
\end{eqnarray}
where $\epsilon(\boldsymbol{S}_i)$ is the delocalization parameter determined by the dynamics of the quark system rather than free parameters. In this way, the system can choose its most favorable configuration through its dynamics in a larger Hilbert space.

\begin{table}[t]
\caption{The Masses (in unit of MeV) of the ground mesons. The experimental values are taken
from the Particle Data Group (PDG)~\cite{ParticleDataGroup:2018ovx}.}
\begin{tabular}{p{1.5cm}<\centering p{1.25cm}<\centering p{1.25cm}<\centering p{1.25cm}<\centering p{1.25cm}<\centering p{1.25cm}<\centering}
\toprule[1pt]
&$K$  &$K^{*}$ &$B_{s}$ &$B_{s}^{*}$  \\
\midrule[1pt]
Expt  &495      & 892   &5366         &5415             \\
ChQM  &495      & 892   &5367         &5415            \\
QDCSM &495      & 892   &5367         &5415            \\
\midrule[1pt]
&$\eta_{b}$ &$\Upsilon$ &$B$ &$B^{*}$ \\
\midrule[1pt]
Expt  &9398                 &9459         &5280    &5325         \\
ChQM  &9398                 &9459         &5280    &5325       \\
QDCSM &9398                 &9459         &5280    &5325       \\
\bottomrule[1pt]
\end{tabular}
\label{mass}
\end{table}

Besides the parameters introduced by the screening color confinement in Eq. (\ref{QDCSM-vc}), almost all the parameters in ChQM and QDCSM are the same, which are determined by reproducing the mass of the low lying meson states in Table \ref{mass}. All the parameters relevant to the present estimations are collected in Table \ref{parameters}. The masses of the Goldston boson are taken as the experimental values but in unit of fm$^{-1}$. As for the  parameter $b$, it is indicating the size of the subcluster. However, the color-octet cluster is not the real physical state, so the value of parameter $b$ could only be determined by fitting the size of the mesons. By using Gaussian expansion method~\cite{Hiyama:2005cf}, we estimated the size of the mesons and find a proper values of parameter $b$, which is $0.2$ fm~\cite{GEM}. In the present estimations, the parameter $a_{c}$ is taken as $a_{c}=101$ MeV fm$^{-2}$, which is the same as the one in Ref.~\cite{GEM}.  As for the parameters $\alpha_{s}^{q_{i}q_{i}}$, they are defined by fitting the mass difference between mesons, while the parameters $V_{0_{q_{i}q_{j}}}$  is determined by fitting the mass shift of the absolute and experimental value of each meson. These parameters are related to the flavor and therefore inevitably increase the number of parameters. In addition, in two kinds of models, almost all parameters are identical except for $V_{0_{us}}$.

\renewcommand\arraystretch{1.25}
\begin{table}[ht]
\caption{The values of the Model parameters. All the parameters in both quark model are the same except for $V_{0_{us}}$, the former value in the tablet is adopted in QDCSM, while the later one is the value taken in ChQM.}
\begin{tabular}{p{2.5cm}<\centering p{2.0cm}<\centering p{3.2cm}<\centering }
\toprule[1pt]
 & Parameter  &Value
    \\
\midrule[1pt]
Quark Mass  & $m_{u/d}$ & 313\\
(MeV)       & $m_s$     & 536\\
            & $m_b$     & 5112 \\
\midrule[1pt]
                  & $m_\pi$ & 0.7 \\
Goldstone boson   & $m_K$  & 2.51\\
 mass (fm)        & $m_\eta$  & 2.77\\
                  & $m_\sigma$ &3.42\\
                  & $m_{f_0/a_0/\kappa}$  & 4.97\\
\midrule[1pt]
                     & $\Lambda_{\pi/\sigma}$ &4.2\\
 Cutoff (fm$^{-1}$)  & $\Lambda_{\eta /K}$    &5.2\\
                     & $\Lambda_{f_0/a_0/\kappa}$  &5.2\\

\midrule[1pt]
          & $V_{0_{us}}$   &-3.7467/-3.7298  \\
$V_{0_{q_{i}q_{i}}}$&$V_{0_{ub}}$  &-2.6750  \\
(fm$^{2}$)     & $V_{0_{sb}}$    &-1.7566\\
          & $V_{0_{bb}}$    &2.6857 \\
\midrule[1pt]
& $\alpha_{s}^{us}$  &0.0716 \\
$\alpha_{s}^{q_{i}q_{i}}$ & $\alpha_{s}^{ub}$ &0.1057 \\
& $\alpha_{s}^{sb}$   &0.1930 \\
& $\alpha_{s}^{bb}$   &2.3401 \\
\midrule[1pt]
& $b$ (fm)  & 0.2 \\
&$a_c$ (MeV fm$^{-2}$)&  101 \\
&$\frac{g_{ch}^{2}}{4 \pi}$ &0.54\\
&$\theta_{p}$  & -15 \\
\bottomrule[1pt]
\end{tabular}
\label{parameters}
\end{table}

\subsection{The wave function}
\begin{figure}[!htb]
\includegraphics[scale=0.55]{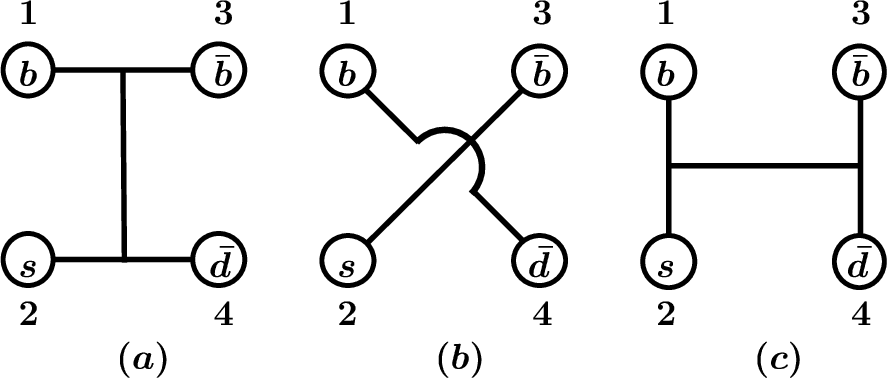}
\vspace{0.5cm} \caption{Two types of configurations in $b\bar b s\bar d$ tetraquarks system. diagram (a) and  (b) are the meson-meson configurations while diagram (c) is diquark-antidiquark configuration. }
\label{fig1}
\end{figure}

In the present work, we focus on the strange hidden bottom system by using the resonance group method~\cite{Kamimura:1981oxj}.
In Fig.~\ref{fig1}, we present two typical kinds of configurations of this system, which are the meson-meson structures as shown in diagrams Fig.~\ref{fig1}-(a) and (b), and the diquark-antidiquark structure as shown in Fig.~\ref{fig1}-(c).

 To solve such a 4-body problem, currently, an economic way is used to combine these two configurations to see the effect of the multi-channel coupling. Four fundamental degrees of freedom, which are color, spin, flavor, and orbit are generally accepted by the QCD theory at the quark level. The multiquark system's wave function is an internal product of the color, spin, flavor, and orbit terms.

\subsubsection{The color wave function}
comparing to the conventional hadrons such as  $q\bar{q}$ mesons and $qqq$ baryons, there are more complicate color structures in multiquark systems. In the following, we construct the colorless wave function of the tetraquark system step by step.

For the meson-meson configurations, the color wave functions of a $q\bar{q}$ cluster can be,
\begin{eqnarray}
C^{1}_{[111]} &=& \sqrt{\frac{1}{3}}(r\bar{r}+g\bar{g}+b\bar{b}), \nonumber\\
C^{2}_{[21]} &=& r\bar{b},\quad   C^{3}_{[21]} =  -r\bar{g},                     \nonumber\\
C^{4}_{[21]} &=& g\bar{b},\quad  C^{5}_{[21]} =  -b\bar{g},               \nonumber\\
C^{6}_{[21]} &=& g\bar{r}, \quad C^{7}_{[21]} =   b\bar{r},          \nonumber\\
C^{8}_{[21]} &=&  \sqrt{\frac{1}{2}}(r\bar{r}-g\bar{g}),        \nonumber\\
C^{9}_{[21]} &=&  \sqrt{\frac{1}{6}}(-r\bar{r}-g\bar{g}+2b\bar{b}),
\end{eqnarray}

where the subscripts $[111]$ and $[21]$ stand for color-singlet ($\textbf{1}_{c}$) and color-octet ($\textbf{8}_{c}$), respectively. The wave functions of the color-singlet can be constructed by the product of two color-singlet cluters, $\textbf{1}_{c}\otimes\textbf{1}_{c}$ (named color-singlet channel) or the product of two color-octet clusters, $\textbf{8}_{c}\otimes\textbf{8}_{c}$ (named hidden-color channel), then the color wave function of the tetraquark system with meson-meson configuration reads,
\begin{equation}
\chi^{c}_{1} = C^{1}_{[111]}C^{1}_{[111]},
\end{equation}
\begin{equation}
\begin{split}
  \chi^{c}_{2} =&\sqrt{\frac{1}{8}}(C^{2}_{[21]}C^{7}_{[21]}-C^{4}_{[21]}C^{5}_{[21]}-C^{3}_{[21]}C^{6}_{[21]}\\
                &+C^{8}_{[21]}C^{8}_{[21]}-C^{6}_{[21]}C^{3}_{[21]}+C^{9}_{[21]}C^{9}_{[21]}\\
                &-C^{5}_{[21]}C^{4}_{[21]}+C^{7}_{[21]}C^{2}_{[21]}).
\end{split}
\end{equation}

For the diquark-antidiquark configuration, the color wave functions of the diquark clusters can be,
\begin{eqnarray}
\nonumber
  C^{1}_{[2]} &=& rr, \quad C^{2}_{[2]} = \sqrt{\frac{1}{2}}(rg+gr), \\ \nonumber
  C^{3}_{[2]} &=& gg, \quad C^{4}_{[2]} = \sqrt{\frac{1}{2}}(rb+br),\\ \nonumber
  C^{5}_{[2]} &=& \sqrt{\frac{1}{2}}(gb+bg), \quad C^{6}_{[2]} = bb, \\ \nonumber
  C^{7}_{[11]} &=& \sqrt{\frac{1}{2}}(rg-gr),\quad  C^{8}_{[11]} = \sqrt{\frac{1}{2}}(rb-br),\\ \nonumber
  C^{9}_{[11]} &=&\sqrt{\frac{1}{2}}(gb-bg).
\end{eqnarray}
While the color wave functions of the antidiquark clusters read,
\begin{eqnarray}
\nonumber
  C^{1}_{[22]} &=& \bar{r}\bar{r}, \quad  C^{2}_{[22]} = -\sqrt{\frac{1}{2}}(\bar{r}\bar{g}+\bar{g}\bar{r}),\\  \nonumber
  C^{3}_{[22]} &=& \bar{g}\bar{g}, \quad  C^{4}_{[22]} = \sqrt{\frac{1}{2}}(\bar{r}\bar{b}+\bar{b}\bar{r}), \\  \nonumber
  C^{5}_{[22]} &=& -\sqrt{\frac{1}{2}}(\bar{g}\bar{b}+\bar{b}\bar{g}), C^{6}_{[22]} = \bar{b}\bar{b}, \\ \nonumber
  C^{7}_{[211]}&=& \sqrt{\frac{1}{2}}(\bar{r}\bar{g}-\bar{g}\bar{r}), \quad  C^{8}_{[211]} = -\sqrt{\frac{1}{2}}(\bar{r}\bar{b}-\bar{b}\bar{r}), \\ \nonumber
  C^{9}_{[211]} &=& \sqrt{\frac{1}{2}}(\bar{g}\bar{b}-\bar{b}\bar{g}). \\
\end{eqnarray}
The color-singlet wave functions of the diquark-antidiquark configuration can be the product of color sextet and antisextet clusters ($\textbf{6}_{c}\otimes\bar{\textbf{6}}_{c}$) or  the product of color-triplet and antitriplet clusters ($\bar{\textbf{3}}_{c} \otimes \textbf{3}_{c}$), which read,
\begin{equation}
\begin{split}
\chi^{c}_{3} = &\sqrt{\frac{1}{6}}(C^{1}_{[2]}C^{1}_{[22]}-C^{2}_{[2]}C^{[2]}_{[22]}+C^{3}_{[2]}C^{3}_{[22]} \\
               &+C^{4}_{[2]}C^{4}_{[22]}-C^{5}_{[2]}C^{5}_{[22]}+C^{6}_{2}C^{6}_{22}),
\end{split}
\end{equation}

\begin{equation}
\begin{split}
  \chi^{c}_{4} =&\sqrt{\frac{1}{3}}(C^{7}_{[11]}C^{7}_{[211]}-C^{8}_{[11]}C^{8}_{[211]}+C^{9}_{[11]}C^{9}_{[211]}).
\end{split}
\end{equation}

\subsubsection{The flavor wave function}
For the flavor degree of freedom, since the quark content of the investigated tetraquark system is $b\bar{b}s\bar{q}$ ($q=(u,d)$), the isospin of this tetraquark system is $I=\frac{1}{2}$. Here, we adopt $F^{i}_{m}$ and $F^{i}_{d}$ to denote the flavor wave functions of the tetraquark system with meson-meson and diquark-antidiquark configurations, respectively. For the meson-meson configuration, two different coupling orders can be accessed, which are
\begin{eqnarray}
  \nonumber  F^{1}_{m}&=& (b\bar{b})(s\bar{q}),   \qquad F^{2}_{m}= (b\bar{q})(s\bar{b}).  \\
\end{eqnarray}
For the diquark-antidiquark configuration, there is only possible flavor wave function, which is
\begin{eqnarray}
  F^{3}_{d}&=& (bs)(\bar{b}\bar{q}).
\end{eqnarray}

\subsubsection{The spin wave function}
For the spin part, the total spin $S$ of tetraquark system can be 0, 1, 2, which can be constructed by the spin of the two body cluster. The spin wave functions of two body clusters could be
\begin{eqnarray}
\nonumber \chi_{11}&=& \alpha\alpha,\\
\nonumber \chi_{10} &=& \sqrt{\frac{1}{2}}(\alpha\beta+\beta\alpha),\\
\nonumber \chi_{1-1} &=& \beta\beta ,\\
            \chi_{00} &=& \sqrt{\frac{1}{2}}(\alpha\beta-\beta\alpha).
\end{eqnarray}

Then, the total spin wave functions $S^{i}_{s}$ are obtained by considering the coupling of two subcluster spin wave functions with SU(2) algebra, and the total spin wave functions of tetraquark system can be read as
\begin{eqnarray}
\nonumber S^{1}_{0}&=&\chi_{00}\chi_{00},\\
\nonumber S^{2}_{0}&=&\sqrt{\frac{1}{3}}(\chi_{11}\chi_{1-1}-\chi_{10}\chi_{10}+\chi_{1-1}\chi_{11}),\\
\nonumber S^{3}_{1}&=&\chi_{00}\chi_{11},\\
\nonumber S^{4}_{1}&=&\chi_{11}\chi_{00},\\
\nonumber S^{5}_{1}&=&\sqrt{\frac{1}{2}}(\chi_{11}\chi_{10}-\chi_{10}\chi_{11}), \\
S^{6}_{2}&=&\chi_{11}\chi_{11}.
\end{eqnarray}

\begin{table*}[!htb]
\begin{center}
\caption{\label{channels} The relevant channels for all possible states with different $J^P$ quantum numbers}
\begin{tabular}{p{1.0cm}<\centering p{1.75cm}<\centering p{1.75cm}<\centering p{1.0cm}<\centering p{1.75cm}<\centering p{1.75cm}<\centering p{1.0cm}<\centering p{1.75cm}<\centering p{1.75cm}<\centering}
\toprule[1pt]
\multicolumn{3}{c}{$I(J^{P})=\frac{1}{2}(0^{+})$} &\multicolumn{3}{c}{$I(J^{P})=\frac{1}{2}(1^{+})$} &\multicolumn{3}{c}{$I(J^{P})=\frac{1}{2}(2^{+})$} \\
  Index    &$F^{i}; S^{j}_{s}; \chi^{c}_{k}$   &Channels      & Index   &$F^{i}; S^{j}_{s}; \chi^{c}_{k}$   &Channels   & Index   &$F^{i}; S^{j}_{s}; \chi^{c}_{k}$    &Channels     \\
\multicolumn{1}{c}{} &[i,j,k] & \multicolumn{1}{c}{}  &\multicolumn{1}{c}{} &[i,j,k] & \multicolumn{1}{c}{} &\multicolumn{1}{c}{} &[i,j,k] & \multicolumn{1}{c}{} \\
\midrule[1pt]
 1 & [2,1,1] &$B_{s}\bar{B}$           &1 & [2,3,1] &$B_{s}\bar{B}^{*}$         &1 & [2,6,1] &$B_{s}^{*}\bar{B}^{*}$\\
 2 & [2,2,1] &$B_{s}^{*}\bar{B}^{*}$   &2 & [2,4,1] &$B_{s}^{*}\bar{B}$         &2 & [1,6,1] &$\Upsilon \bar{K}^{*}$\\
 3 & [1,1,1] &$\eta_{b}\bar{K}$        &3 & [2,5,1] &$B_{s}^{*}\bar{B}^{*}$     &3 & [2,6,2] &$B_{s(8)}\bar{B}^{*}_{(8)}$\\
 4 & [1,2,1] &$\Upsilon \bar{K}^{*}$   &4 & [1,3,1] &$\eta_{b}\bar{K}^{*}$      &4 & [1,6,2] &$\Upsilon_{(8)} \bar{K}^{*}_{(8)}$\\
 5 & [2,1,2] &$B_{s(8)}\bar{B}_{(8)}$  &5 & [1,4,1] &$\Upsilon \bar{K}$         &5 & [3,6,3] &$(bs)(\bar{b}\bar{q})$\\
 6 & [2,2,2] &$B_{s(8)}^{*}\bar{B}^{*}_{8}$     &6 & [1,5,1] &$\Upsilon \bar{K}^{*}$       &6 & [3,6,4] &$(bs)(\bar{b}\bar{q})$\\
 7 & [1,1,2] &$\eta_{b(8)}\bar{K}_{(8)}$        &7 & [2,3,2] &$B_{s(8)}\bar{B}^{*}_{(8)}$&  &&\\
 8 & [1,2,2] &$\Upsilon_{(8)}\bar{K}^{*}_{(8)}$ &8 & [2,4,2] &$B_{s(8)}^{*}\bar{B}_{(8)}$    &  && \\
 9 & [3,1,3] &$(bs)(\bar{b}\bar{q})$ &9 & [2,5,2] &$B_{s(8)}^{*}\bar{B}^{*}_{(8)}$ &  && \\
 10& [3,2,4] &$(bs)(\bar{b}\bar{q})$ &10 & [1,3,2] &$\eta_{b(8)}\bar{K}^{*}_{(8)}$ &  &&\\
 11& [3,1,3] &$(bs)(\bar{b}\bar{q})$ &11 & [1,4,2] &$\Upsilon_{(8)} \bar{K}_{(8)}$   &  && \\
 12& [3,2,4] &$(bs)(\bar{b}\bar{q})$ &12 & [1,5,2] &$\Upsilon_{(8)} \bar{K}^{*}_{(8)}$ & &&\\
   &&         &13 & [3,3,3]&$(bs)(\bar{b}\bar{q})$  &  &&\\
   &&         &14& [3,3,4] &$(bs)(\bar{b}\bar{q})$  &  &&\\
   &&         &15& [3,4,3] &$(bs)(\bar{b}\bar{q})$ &   &&\\
   &&         &16& [3,4,4] &$(bs)(\bar{b}\bar{q})$ &   &&\\
   &&         &17& [3,5,3] &$(bs)(\bar{b}\bar{q})$ &   &&\\
   &&         &18& [3,5,4] &$(bs)(\bar{b}\bar{q})$ &   &&\\
\bottomrule[1pt]
\end{tabular}
\end{center}
\label{channe}
\end{table*}

\subsubsection{The orbital wave function}
In the spatial space, we define different Jacobi coordinates for different diagrams in Fig. \ref{fig1}. As for the meson-meson configuration in digram (a), the Jacobi coordinates are
\begin{eqnarray}
   \R_{1}&=&\r_{1}-\r_{2},\quad \R_{2}=\r_{3}-\r_{4},  \nonumber \\
             \R &=&\frac{\r_{1}+\r_{2}}{2}-\frac{\r_{3}+\r_{4}}{2}
\end{eqnarray}
By interchanging $\r_2$ with $\r_4$ one can obtain the Jacobi coordinates for diagram (b) in Fig.~\ref{fig1}. As for the diquark-antidiquark configuration as shown in  Fig.~\ref{fig1}-(c), the Jacobi coordinates are defined as,
\begin{eqnarray}
   \R_{1}&=&\r_{1}-\r_{3},\quad \R_{2}=\r_{2}-\r_{4}, \nonumber  \\
             \R &=&\frac{\r_{1}+\r_{3}}{2}-\frac{\r_{2}+\r_{4}}{2}.
\end{eqnarray}
With Jacobi coordinates defined above, we use the resonating group method (RGM) to solve the Schr\"{o}dinger-like 4-body bound state equation~\cite{Kamimura:1981oxj}. The total orbital wave functions can be constructed by the product of the orbital wave functions of two internal clusters and the relative motion wave function between the two clusters, which is,
\begin{equation}
\psi^{L}_{4q}=\psi_{1}(\R_{1})\psi_{2}(\R_{2})\chi_{L}(\R),\label{wave1}
\end{equation}
where $\R_{1}$ and $\R_{2}$ are the internal Jacobi coordinates of cluster 1 and cluster 2, respectively. $\R$ is the relative coordinate between the two clusters 1 and 2. $\psi_{1}(\R_1)$ and $\psi_{2}(\R_2)$ are the internal cluster orbital wave functions of the clusters 1 and clusters 2, respectively, while $\chi_{L}(\R)$ is the relative motion wave function between two clusters, which can be expanded by the gaussian bases
\begin{eqnarray}
\begin{split}
\chi_{L}(\R)=&\sqrt{\frac{1}{4\pi}}\left(\frac{1}{\pi b^2}\right)^{3/4}\sum^{n}_{i=1}C_{i,L} \\
   &\times \int{\mathrm{exp}}\left[-\frac{1}{2b^2}\big(\R-\boldsymbol{S_{i}}\big)^2\right]Y_{LM}(\hat{\boldsymbol{S_{i}}})d\hat{s_{i}},
\end{split}
\end{eqnarray}
where $C_{i,L}$ is the expansion coefficient, and $n$ is the number of gaussian bases, which is determined by the stability of the results. By including the center of mass motion,
\begin{equation}
  \phi_{C}(\boldsymbol{R_{c}})=(\frac{4}{\pi b^2})^{3/4}\mathrm{e}^{\frac{-2 \boldsymbol{R_{c}}^{2}}{b^{2}}},
\end{equation}
the ansatz, Eq.~(\ref{wave1}), can be rewritten as
\begin{eqnarray}
\begin{split}
\psi^{L}_{4q}=\sum_{i=1}^{n}C_{i, L}\int \frac{d \hat{\boldsymbol{S_{i}}}}{\sqrt{4 \pi}} \prod_{\alpha=1}^{2}\phi_{\alpha}(\boldsymbol{S_{i}})\prod_{\alpha=3}^{4}\phi_{\beta}(\boldsymbol{-S_{i}}) ,\\
\end{split}
\end{eqnarray}
where $\phi_{\alpha}(\boldsymbol{S_{i}})$ and $\phi_{\beta}(\boldsymbol{-S_{i}})$ are the single-particle orbital wave functions, their specific form are shown in Eq.~(\ref{wave2}).

Finally, to fulfill the Pauli principle, the complete wave function is written as
 \begin{equation}
   \psi_{4q}=\mathcal{A}\left[\left[\psi^{L}_{4q}S^{j}_{s}\right]_{JM_{J}}F^{i}_{I}\chi^{c}_{k}\right],
 \end{equation}
where $A$ is the antisymmetry operator of double-heavy tetraquarks. In the present work, the operator $A$ is the unit operator due to the absence of any homogeneous quarks in the strange hidden bottom system.

\section{RESULTS AND DISCUSSIONS}{\label{results}}

In the present calculation, we only investigate all the possible $S$-wave hidden bottom tetraquark with strangeness by taking into account the meson-meson and diquark-antidiquark configurations in the ChQM and QDCSM. Accordingly, the total orbital angular momentum $L$ is 0 and the parity of tetraquark states are positive. The total angular momentum $J$ coincides with the total spin, which could be 0, 1, 2, respectively. For the considered $b\bar{b} s\bar{q}$ system, the isospin is $1/2$. All the possible states with different $J^P$ quantum numbers are listed in Table~\ref{channels}, where $F^{i}, S_{s}^{j}, \chi_{k}^{c}$ correspond to the degrees of freedom of spin, flavor, and color, respectively. For the meson-meson structure, both the color singlet-singlet $( \textbf{1}_{c}\otimes\textbf{1}_{c})$ and the color octet-octet $(\textbf{8}_{c}\otimes \textbf{8}_{c})$ are taken into account in ChQM, whereas since the introduction of the delocalization parameter and color screening in QDCSM which have the same effect of hidden color channel coupling~\cite{Huang:2011kf} to some extent, only the color singlet-singlet $(\textbf{1}_{c}\otimes \textbf{1}_{c})$ is calculated in this model. For the diquark-antidiquark, two color configurations, antitriplet-triplet $(\bar{\textbf{3}}_{c} \otimes \textbf{3}_{c})$ and sextet-antisextet $(\textbf{6}_{c}\otimes\bar{\textbf{6}}_{c})$ are considered in both models


\begin{table*}[!htb]
\begin{center}
\caption{\label{0} The lowest-lying eigenenergies of the strange hidden bottom tetraquarks with $I(J^{P})=\frac{1}{2}(0^+)$ in the ChQM and QDCSM. }
\begin{tabular}{p{1.0cm}<\centering p{1.5cm}<\centering p{1.75cm}<\centering p{1.2cm}<\centering p{1.2cm}<\centering p{1.2cm}<\centering p{1.2cm}<\centering p{1.2cm}<\centering p{1.2cm}<\centering p{1.2cm}<\centering}
\toprule[1pt]
\multirow{2}{*}{Index}  & \multirow{2}{*}{Channel}   & \multirow{2}{*}{Threshold} & \multicolumn{3}{c}{ChQM} &  & \multicolumn{3}{c}{QDCSM} \\
\cmidrule[1pt]{4-6} \cmidrule[1pt]{8-10}
   &    &   &$E_{sc}$  &$E_{cc}$  &$E_{mix}$ & &$E_{sc}$  &$E_{cc}$  &$E_{mix}$\\
\midrule[1pt]
1 &$B_{s}\bar{B}$                 & 10647   &10649   & 9897   &9897      & &10547    &9897      &9897\\
2 &$B_{s}^{*}\bar{B}^{*}$         & 10740   &10742   &        &          & &10597    &          &\\
3 &$\eta_{b}\bar{K}$              & 9894    &9897    &        &          & &9897     &          &\\
4 &$\Upsilon \bar{K}^{*}$         & 10352   &10354   &        &          & &10354    &          &\\
5 &$B_{s(8)}\bar{B}_{(8)}$        &         &11242   &11083   &          & &         &          &\\
6 &$B_{s(8)}^{*}\bar{B}^{*}_{8}$  &         &11171   &        &          & &         &          &\\
7 &$\eta_{b(8)}\bar{K}_{(8)}$     &         &11123   &        &          & &         &          &\\
8 &$\Upsilon_{(8)}\bar{K}^{*}_{(8)}$&       &11171   &        &          & &         &          &\\
9&$(bs)(\bar{b}\bar{q})$          &         &11090   & 10960  &          & &10648    &10521     &\\
10&$(bs)(\bar{b}\bar{q})$         &         &11248   &        &          & &10710    &          &\\
11&$(bs)(\bar{b}\bar{q})$         &         &11100   &        &          & &10618    &          &\\
12&$(bs)(\bar{b}\bar{q})$         &         &11128   &        &          & &10595    &          &\\
\bottomrule[1pt]
\end{tabular}
\end{center}
\end{table*}

\begin{table*}[!htb]
\begin{center}
\caption{\label{1} The same as Table \ref{0} but for tetraquarks with $I(J^P)=\frac{1}{2}(1^+)$ states. }
\begin{tabular}{p{1.0cm}<\centering p{1.5cm}<\centering p{1.75cm}<\centering p{1.2cm}<\centering p{1.2cm}<\centering p{1.2cm}<\centering p{1.2cm}<\centering p{1.2cm}<\centering p{1.2cm}<\centering p{1.2cm}<\centering}
\toprule[1pt]
\multirow{2}{*}{Index}  & \multirow{2}{*}{Channel}   & \multirow{2}{*}{Threshold} & \multicolumn{3}{c}{ChQM} &  & \multicolumn{3}{c}{QDCSM} \\
\cmidrule[1pt]{4-6} \cmidrule[1pt]{8-10}
   &    &   &$E_{sc}$  &$E_{cc}$  &$E_{mix}$ & &$E_{sc}$  &$E_{cc}$  &$E_{mix}$\\
\midrule[1pt]
1 &$B_{s}\bar{B}^{*}$      & 10692         &10694      & 9958      &9958      &    &10576    &9958      &9958\\
2 &$B_{s}^{*}\bar{B}$      & 10695         &10697      &           &          &    &10578    &          &\\
3 &$B_{s}^{*}\bar{B}^{*}$  & 10740         &10742      &           &          &    &10596    &          &\\
4 &$\eta_{b}\bar{K}^{*}$   & 10291         &10293      &           &          &    &10293    &          &\\
5 &$\Upsilon \bar{K}$      & 9955          &9958       &           &          &    &9958     &          &\\
6 &$\Upsilon \bar{K}^{*}$  & 10352         &10354      &           &          &    &10354    &          &\\
7 &$B_{s(8)}\bar{B}^{*}_{(8)}$       &     &11236      & 11036     &          &    &         &          &\\
8 &$B_{s(8)}^{*}\bar{B}_{(8)}$       &     &11236      &           &          &    &         &          &\\
9 &$B_{s(8)}^{*}\bar{B}_{(8)}^{*}$   &     &11171      &           &          &    &         &          &\\
10&$\eta_{b(8)}\bar{K}^{*}_{(8)}$    &     &11116      &           &          &    &         &          &\\
11&$\Upsilon_{(8)} \bar{K}_{(8)}$    &     &11193      &           &          &    &         &          &\\
12&$\Upsilon_{(8)} \bar{K}^{*}_{(8)}$  &   &11178      &           &          &    &         &          &\\
13&$(bs)(\bar{b}\bar{q})$              &   &11112      & 10974     &          &    &10660    &10560     &\\
14&$(bs)(\bar{b}\bar{q})$              &   &11237      &           &          &    &10704    &          &\\
15&$(bs)(\bar{b}\bar{q})$              &   &11114      &           &          &    &10661    &          &\\
16&$(bs)(\bar{b}\bar{q})$              &   &11236      &           &          &    &10704    &          &\\
17&$(bs)(\bar{b}\bar{q})$              &   &11118      &           &          &    &10645    &          &\\
18&$(bs)(\bar{b}\bar{q})$              &   &11177      &           &          &    &10646    &          &\\
\bottomrule[1pt]
\end{tabular}
\end{center}
\end{table*}

\begin{table*}[!htb]
\begin{center}
\caption{\label{2} The same as Table \ref{0} but for tetraquarks with $I(J^P)=\frac{1}{2}(2^+)$ states. }
\begin{tabular}{p{1.0cm}<\centering p{1.5cm}<\centering p{1.75cm}<\centering p{1.2cm}<\centering p{1.2cm}<\centering p{1.2cm}<\centering p{1.2cm}<\centering p{1.2cm}<\centering p{1.2cm}<\centering p{1.2cm}<\centering}
\toprule[1pt]
\multirow{2}{*}{Index}  & \multirow{2}{*}{Channel}   & \multirow{2}{*}{Threshold} & \multicolumn{3}{c}{ChQM} &  & \multicolumn{3}{c}{QDCSM} \\
\cmidrule[1pt]{4-6} \cmidrule[1pt]{8-10}
   &    &   &$E_{sc}$  &$E_{cc}$  &$E_{mix}$ & &$E_{sc}$  &$E_{cc}$  &$E_{mix}$\\
\midrule[1pt]
1    &$B_{s}^{*}\bar{B}^{*}$              & 10740   &10742   &10354      & 10354    &    &10617    &10354     &10354\\
2    &$\Upsilon \bar{K}^{*}$              & 10352   &10354   &           &          &    &10354    &          &\\
3    &$B_{s(8)}^{*}\bar{B}^{*}_{8}$       &         &11287   &11148      &          &    &         &          &\\
4    &$\Upsilon_{(8)}\bar{K}^{*}_{(8)}$   &         &11193   &           &          &    &         &          &\\
5    &$(bs)(\bar{b}\bar{q})$              &         &11153   &11001      &          &    &10699    &10635     &\\
6    &$(bs)(\bar{b}\bar{q})$              &         &11271   &           &          &    &10749    &          &\\
\bottomrule[1pt]
\end{tabular}
\end{center}
\end{table*}

\subsection{Possible bound states}

In Tables~\ref{0}-\ref{2}, we list our estimations of the lowest-lying strange hidden bottom tetraquark states with quantum numbers $I(J^{P})=\frac{1}{2}(0^{+}), \frac{1}{2}(1^{+}), \frac{1}{2}(2^{+})$, respectively. In these tables, all the allowed meson-meson and diquark-antidiquark configurations are listed in the 2nd column and the experimental values of the thresholds for the physical channels are presented in the 3rd column. In the table, $E_{sc}$, $E_{cc}$, and $E_{mix}$ are the eigen of every single channel, the eigen of the couple channels for each kind of configurations, and the estimated eigen by simultaneously considering the meson-meson and diquark-antidiquark configurations, respectively.

\subsubsection{ChQM estimations}
The eigenenergies of the relevant strange hidden-bottom tetraquak states are listed in Tables~\ref{0}-\ref{2}. The estimations in ChQM indicate that the eigenenergies of each single channel for meson-meson structures are higher than the corresponding thresholds, which demonstrates that the formation of a bound state is impossible. For the diquark-antidiquark structures, the eigenenergies of each single channel are even a bit larger than those of the single-channel energies of the meson-meson structures. Then, we consider the coupling of the color configurations $(\textbf{1}_{c}\otimes \textbf{1}_{c})$ and $(\textbf{8}_{c}\otimes  \textbf{8}_{c})$ individually in meson-meson structures, and coupled channels estimations indicate that the lowest energies for considered $0^+$, $1^+$ and $2^+$ states are still above the corresponding lowest meson-meson threshold. As for the diquark-antiquark configuration, our estimations indicate that the coupling of the different color configurations, i.e., $(\bar{\textbf{3}}_{c} \otimes \textbf{3}_{c})$ and $(\textbf{6}_{c}\otimes \bar{\textbf{6}}_{c})$, are non-negligible. However, even when we including the coupled channels effects in the diquark-antidiquark configuration, the estimated eigenenergies for different $J^P$ quantum numbers are still above the corresponding lowest meson-meson thresholds. Finally, we perform a completed coupled channels estimations by involving the mixing between the meson-meson and diquark-antidiquark configurations, and we find the estimated eigenenergies for $0^+$, $1^+$ and $2^+$ states are $9897$, $9958$ and $10354$ MeV, respectively, which are still above the corresponding lowest meson-meson thresholds.

To summarize, the present estimations in the ChQM indicate that there are not bound strange hidden-bottom tetraquark states below the lowest meson-meson threshold. However, it should be noticed that due to the color structure of the hidden color channels, there may be resonance states above the meson-meson thresholds, which will be further evaluated by using the stabilization methods in the following subsection.

\subsubsection{QDCSM estimations}
In the present work, we also investigate the strange hidden-bottom tetraqaurk system in the QDCSM, and the results are also listed in Tables~\ref{0}-\ref{2}. Different from the estimations in ChQM, only the color singlet-singlet structure i.e., $(\textbf{1}_{c}\otimes\textbf{1}_{c})$ is taken into account in the meson-meson configuration, because the phenomenological color screening confinement is effective description of the hidden color channel coupling~\cite{Huang:2011kf} in QDCSM.

As for the $I(J^{P})=\frac{1}{2}(0^{+})$ system, four physical channels in meson-meson structure and four diquark-antiquark channels are explored in QDCSM. As shown in the Table~\ref{0}, the single channel estimations indicate that the eigenenergies of $B_{s}\bar{B}$ and $B_{s}^{*}\bar{B}^{*}$ systems are both below the corresponding thresholds, with the binding energy of 100 MeV and 143 MeV, respectively. As for the other two channels, i.e., $\eta_{b}\bar{K}$, $\Upsilon \bar{K}^{*}$, the estimated eigenenergies are above the corresponding thresholds. For the diquark-antidiquark configuration, all the estimated masses are higher than the lowest meson-meson threshold in the current single channel estimations, and the minimum energy is 10595 MeV. The coupled channels estimations show that eigenenergies of strange hidden-bottom tetraquark states are 9897 and 10521 MeV for the meson-meson and diquark-antidiquark structures, respectively. The eigenenergy for the meson-meson structure is almost the same as the lowest meson-meson threshold, which indicate that no bound states can be formed in the meson-meson structure. For the diquark-antidiquark structure, although the eigenenergies become lower compared to the eigenenergies of singel channel estimations, the lowest energy is still higher than the lowest threshold of the meson-meson channel. When considering the mixing of the meson-meson and  diquark-antidiquark structures, the estimated eigenenergy illustrates that the bound state is still unavailable by comparing the lowest $\eta_{b}\bar{K}$ threshold. Nevertheless, it is most possible to make resonance states for the $B_{s}\bar{B}$ and $B_{s}^{*}\bar{B}^{*}$ by the channel coupling, which will be discussed in the following subsection.

For the $I(J^{P})=\frac{1}{2}(1^{+})$ system, the behavior is similar to that of the $I(J^{P})=\frac{1}{2}(0^+)$ system. There are 12 channels in this case which include six color singlet meson-meson configurations and six channels in the diquark-antidiquark arrangement. The single-channel calculation shows that there are only three physical channels, $B_{s}\bar{B}^*$, $B_{s}^*\bar{B}$, $B_{s}^{*}\bar{B}^{*}$, which can be considered as bound states and the estimated eigenenergies are more than 100 MeV below the corresponding meson-meson thresholds. The single channel estimations indicate that the eigenenergy of each diquark-antidiquark channel is about $10.5\sim 10.7$ GeV. considering the coupling between channels in the same configurations, the eigneenergies are estimated to be 9958 and 10560 MeV for the meson-meson and diquark-antidiquark configurations, respectively. Further complete coupled-channels calculations predict a state with a mass 9958 MeV, which is still above the threshold of $\Upsilon\bar{K}$.

As for the $I(J^{P})=\frac{1}{2}(2^{+})$ system, there are two meson-meson channels, i.e., $B_{s}^{*}\bar{B}^{*}$, $\Upsilon\bar{K}^*$ and two diquark-antidiquark channels with the color configurations to be $\bar{\textbf{3}}_{c} \otimes \textbf{3}_{c}$ and $\textbf{6}_{c}\otimes \bar{\textbf{6}}_{c}$, respectively. For the meson-meson structure, the single-channel calculations indicate that there is a bound state composed of $B_{s}^{*}\bar{B}^{*}$ with the binding energy of 123 MeV. The estimations of the coupled channels in the same configurations find that the eigenenergies are 10354 and 10635 MeV for the meson-meson and diquark-antidiquark configurations, respectively, which are above the threshold of $\Upsilon K^\ast $. The complete coupled-channels calculations indicate that the mass of the strange hidden-bottom tetraquark states with $I(J^{P})=\frac{1}{2}(2^{+})$ is 10354 MeV, which is also above the threshold of $\Upsilon K^\ast $.

\begin{figure*}
\includegraphics[scale=2]{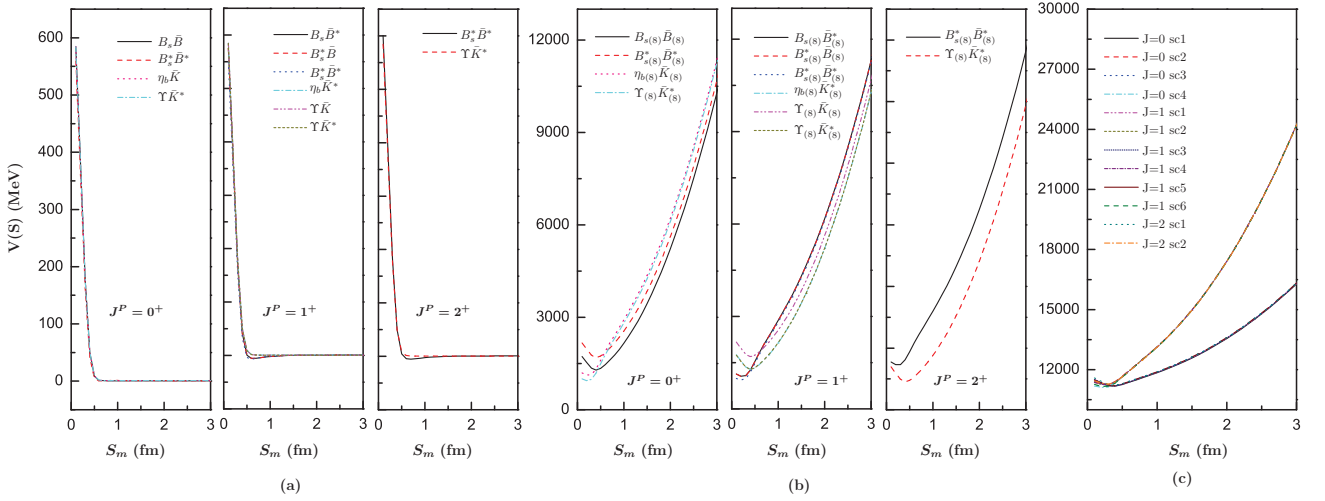}
\caption{The effective potential in the ChQM. Diagram (a) and (b) show the potential of the color singlet and  hidden color channels, respectively, while diagram (c) present the potentials of the diquark-antidiquark channels.  }
\label{Fig:PoChQM}
\end{figure*}

\begin{figure}
\includegraphics[scale=1.1]{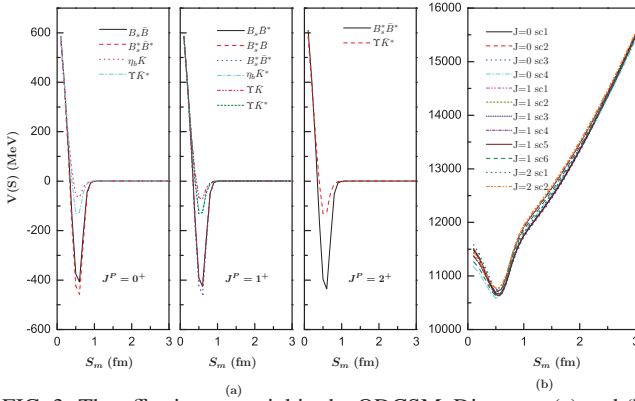}
\vspace{-0.5cm} \caption{The effective potential in the QDCSM. Diagrams (a) and (b) present the potential of color singlet channels in the meson-meson structure and diquark-antidiquark configurations, respectively.  }
\label{Fig:PoQDCSM}
\end{figure}


By Comparing to the results of the two kinds of quark models, one can find the estimated eigenenergies have subtly somewhat divergences and similarities. On the whole, the present estimations indicate that there are no bound state exists for either channel coupling of any configurations or all structures, which are the same for both model. However, the results of the single-channel estimations in QDCSM suggest that several bound states can be formed while there is no any bound state in ChQM. In addition, from Tables~\ref{0}-\ref{2}, one can find that the eigenenergies obtained in QDCSM are generally lower than those in ChQM. To further clarify the reasons of such discrepancies, we have analyzed the effective potential of each channel in the strange hidden bottom system. In Fig. ~\ref{Fig:PoChQM}, we present the relevant effective potentials in the ChQM, where diagrams (a) and (b) are the potential of the color singlet and hidden color channels in the meson-meson configuration, respectively, while diagram (c) are the potentials of the diquark-antidiquark configurations. In Fig.~\ref{Fig:PoQDCSM}, the relevant effective potentials in the QDCSM are presented, where diagrams (a) and (b) present the potential of color singlet channels in the meson-meson configuration and the diquark-antidiquark configuration, respectively. As shown in diagram Fig. \ref{Fig:PoChQM}-(a), the effective potentials of the color single channels for all possible quantum numbers are repulsive in ChQM, and thus, it is impossible to form any bound state in the single channel estimations. As for the QDCSM, it is interesting to notice that $(s\bar{b})$ and $(b\bar{q})$ such meson-meson structures have a deep equivalent attraction, which explains well why bound states appear in single-channel calculations. For the hidden color channels, One can see that in diagram Fig. \ref{Fig:PoChQM}-(b) the minimum effective energy of each channel is approximately at a distance of 0.2 to 0.4 fm. For the diquark-antidiquark channels, because of the color structure, two colorful subclusters can not fall apart, which indicates the energy is increasing in both ChQM and QDCSM when the two subclusters approach closely or fall apart, the resonance states are possible in these configurations.

\begin{figure}
\includegraphics[scale=0.55]{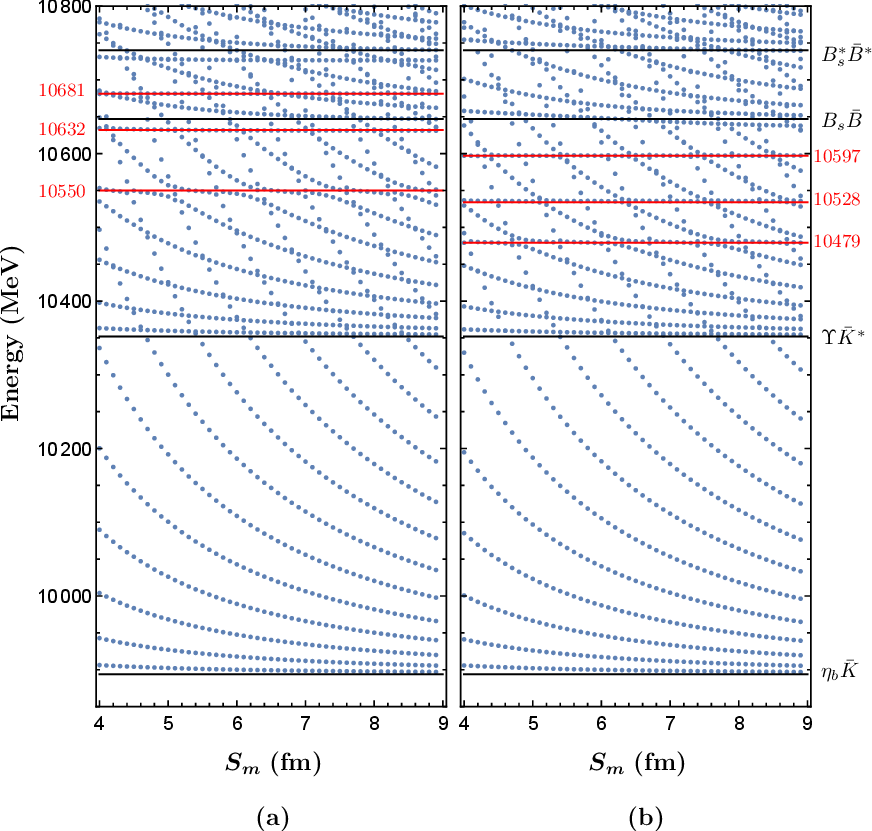}
\caption{(Color online.) The stabilization plots of the energies with $I(J^{P})=\frac{1}{2}(0^+)$ in ChQM (diagram (a)) and QDCSM (diagram (b)).  }
\label{Fig:Stable0}
\end{figure}

\begin{figure}
\includegraphics[scale=0.55]{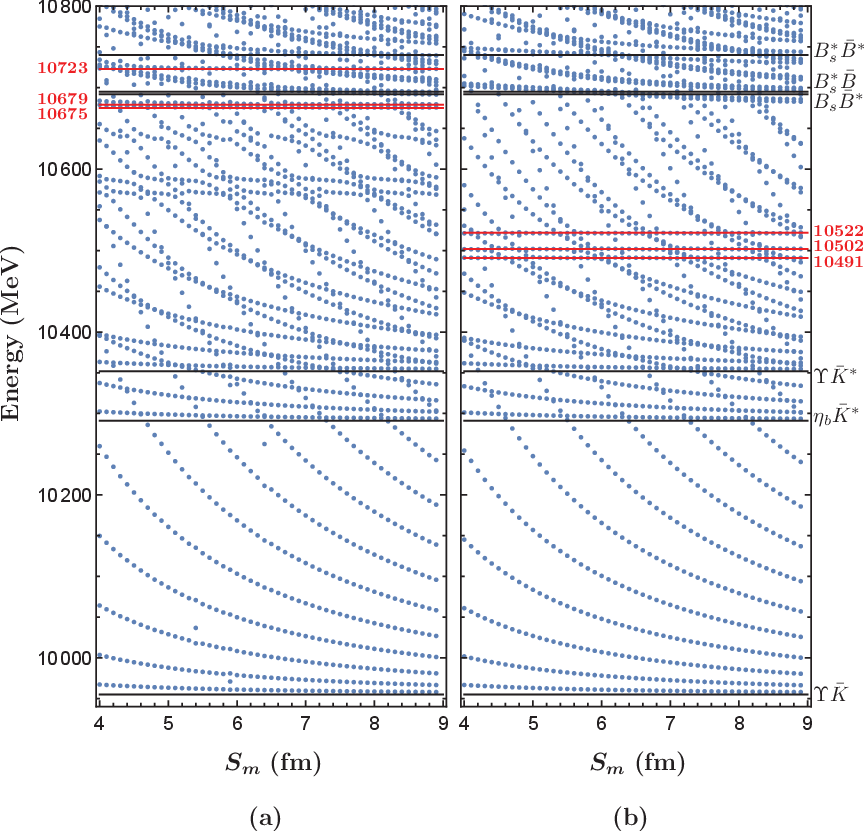}
 \caption{(Color online.) The same as Fig.~\ref{Fig:Stable1} but for $I(J^{P})=\frac{1}{2}(1^+)$ system. }
\label{Fig:Stable1}
\end{figure}

\begin{figure}
\includegraphics[scale=0.55]{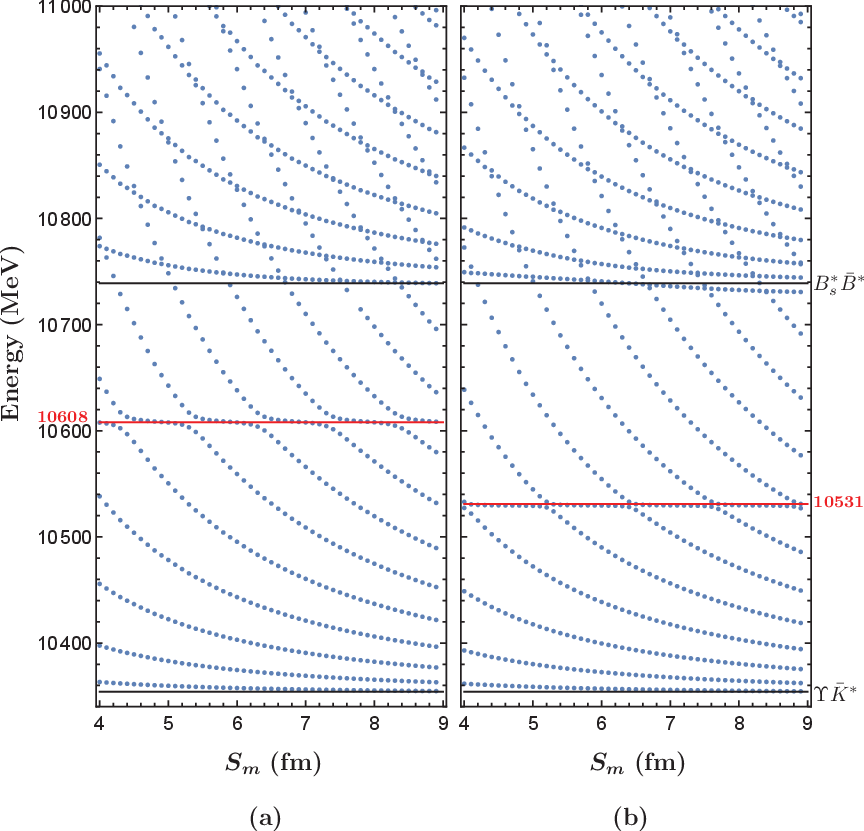}
\caption{(Color online.) The same as Fig.~\ref{Fig:Stable1} but for $I(J^{P})=\frac{1}{2}(2^+)$ system..  }
\label{Fig:Stable2}
\end{figure}

\subsection{Possible resonances}

According to the above analysis,  there may exist several resonance states in the strange hidden bottom system. To further check the resonance possibility in the strange hidden-bottom system, we employ a stabilization method (also named real scaling method), which is a powerful tool to estimate the energies of the metastable states of electron-atom, electron-molecule, and atom-diatom complexes~\cite{real_Method}. And moreover, such a method has also been successfully used in other multiquark systems~\cite{Hiyama:2005cf, Hiyama:2018ukv,real_Method, Jin:2020jfc}, in which some newly observed full heavy tetraquarks states can be well explained. In the present work, the distance between two clusters is defined as $S$, which can  range from 4.0 to 9.0 fm, and with the variation of $S$, an unbound state will fall off toward its threshold, but the resonance states will not be affected by the boundary at a large distance and should be stay stable. Under such criteria, one can find the resonances by checking the stabilization plots of the energies of the strange hidden bottom systems in ChQM and QDCSM. In Figs. \ref{Fig:Stable0}-\ref{Fig:Stable2}, we present stabilization plots of the energies of the strange hidden bottom systems in ChQM and QDCSM with $I(J^P)$ to be $\frac{1}{2}(0^+)$, $\frac{1}{2}(1^+)$ and $\frac{1}{2}(2^+)$, respectively. The black horizontal lines indicate the physical thresholds of the involved meson-meson configurations, while the red horizontal lines refer to the genuine resonances and the concrete masses are marked.

For the $I(J^{P})=\frac{1}{2}(0^{+})$ strange hidden-bottom tetraquark system, our estimation of the energy depending on the distance of the two cluster are presented in Fig. \ref{Fig:Stable0}, where digram (a) and (b) are the estimations in ChQM and QDCSM, respectively. From the figure one can find that there is no resonance between the two lowest meson-meson thresholds for both model. Above the $\Upsilon \bar{K}^\ast$ threshold, both model predict three resonance states, The masses are estimated to be 10550, 10632 and 10681 MeV in the ChQM, while in the QDCSM, the masses are 10479, 10528 and 10597 MeV, which are about 100 MeV below the ones predicted by ChQM. As shown in Fig.~\ref{Fig:Stable1}, we also predicted three resonances in both model, but the estimated masses are rather different, in particular, the masses are evaluated to be 10675, 10679, and 10723 MeV in the ChQM, while those obtained by QDCSM are about 200 MeV below. As for the $I(J^{P})=\frac{1}{2}(2^{+})$ system, there are only two thresholds, which are the thresholds of $\Upsilon \bar{K}^\ast$ and $B_s^\ast \bar{B}^\ast$, respectively. In this system, both model predicted one resonance with a $10608$ and $10531$ MeV fro ChQM and QDCSM, respectively.

\begin{table*}[!htb]
\begin{center}
\caption{\label{resonanceT1} The energies of strange hidden bottom excited states and the percentage of each component for each resonance state in the QDCSM and ChQM.(unit: MeV, "MS" and "MH" stands for the percentage of the meson-meson($\textbf{1}_c\otimes\textbf{1}_c$, $\textbf{8}_c\otimes \textbf{8}_c$)), respectively, while the "DH" represents the percentage of the diquark-antidiquark($\bar{\textbf{3}}_{c} \otimes \textbf{3}_{c}$, $\textbf{6}_{c}\otimes \bar{\textbf{6}}_{c}$)}.
\begin{tabular}{p{1.5cm}<\centering p{5cm}<\centering p{5cm}<\centering p{5cm}<\centering}
\toprule[1pt]
   &$I(J^{P})=\frac{1}{2}(0^{+})$ &$I(J^{P})=\frac{1}{2}(1^{+})$ &$I(J^{P})=\frac{1}{2}(2^{+})$\\
\midrule[1pt]
  \multirow{6}{*}{ChQM} &10681   & 10723     &  10680\\
   ~&(MS:$20\%$, MH:$20\%$, DH:$60\%$)  &(MS:$43\%$, MH:$10\%$, DH:$47\%$)   &(MS:$16\%$, MH:$17\%$, DH:$67\%$)\\
   &10632   & 10679     &\\
   ~&(MS:$29\%$, MH:$28\%$, DH:$43\%$)  &(MS:$33\%$, MH:$25\%$, DH:$42\%$)    \\
   &10550    & 10675     &\\
  &(MS:$21\%$, MH:$16\%$, DH:$63\%$)   &(MS:$43\%$, MH:$13\%$, DH:$44\%$)&\\
 \midrule[1pt]
 \multirow{6}{*}{QDCSM} &10597   & 10522     &  10531\\
   &(MS:$53\%$, DH:$47\%$)     &(MS:$46\%$, DH:$54\%$)   &(MS:$57\%$, DH:$43\%$)\\
   &10528   & 10502     &\\
   &(MS:$50\%$, DH:$50\%$)     &(MS:$47\%$, DH:$53\%$)    \\
   &10479    & 10491     &\\
   &(MS:$63\%$, DH:$37\%$)     &(MS:$37\%$, DH:$63\%$)\\
\bottomrule[1pt]
\end{tabular}
\end{center}
\end{table*}

In Table~\ref{resonanceT1}, we collected the masses of the possible resonances predicted by ChQM and QDCSM. For the $I(J^{P})=\frac{1}{2}(0^{+})$ strange hidden-bottom tetraquark system, three resonances are predicted in both ChQM and QDCSM. Considering the model dependences of the estimations,  the masses of the three resonances are predicted to be  $(10479\sim 10550)$ MeV, $(10528 \sim 10632)$ MeV, and $(10597 \sim 10681)$ MeV, respectively. For the $I(J^{P})=\frac{1}{2}(1^{+})$ system, the masses of the three possible resonances are predicted to be very different, the ChQM predicted masses are around 10500 MeV, while those estimated in QDCSM are in the vicinity of 10700 MeV. In Ref.~\cite{Azizi:2020zyq}, by using the in-medium sum rules, the authors predicted a $Z_{bs}$ tetraquark state with a mass around 10730 MeV. In addition, with the help of the temperature QCD sum rule mode, the authors in Ref~\cite{Sungu:2020zvk} found the mass of $Z_{bs}$ is $M_{Z_{bs}}=10.866^{+0.453}_{-0.403}$ GeV. Combing the above analysis, our predications are consistent with those results within the uncertainty of the model estimations. As for the $I(J^{P})=\frac{1}{2}(2^{+})$ system, only one resonance state can be obtained, the mass is predicted to be $(10531 \sim 10680)$ MeV.

Besides the mass of the possible resonances, we also collect the estimated proportions of each component in Table ~\ref{resonanceT1}. For the ChQM, each color structure has an effect on the resonance state, and the coupling effect of different color structures is very strong, which leads to low resonance energy. From the Table~\ref{resonanceT1}, it can be found that the hidden color channels account for a large proportion of the resonance states, for instance, the hidden color channel component of $10681$ MeV with $I(J^{P})=\frac{1}{2}(0^+)$ is around $80\%$, which indicates that the resonance states prefer to be compact tetraquark states. For the QDCSM, according to the particularity of the model, the meson-meson structure includes color singlet structures as well as other color structures, so from the present results, the color structure has a great influence on the resonance state. Taking the resonance with a mass 10579 MeV as an example, although the proportion of the meson-meson structure is around 53$\%$, here the meson-meson structure also exists some hidden color channel effect, so considering such a situation, the results of the two models are almost similar. According to the analysis of the above results, the color structure coupling not only constitutes the formation mechanism of the resonance state, but the size of the color coupling also has a great influence on the energy of the resonance state~\cite{Hidden}. Based on a systematic analysis of the results, several resonance states are predicted in our present estimation, which may be helpful for future heavy-ion collision experiments and those aiming to study the behavior of the exotic states.

\section{Summary}\label{Sum}

In the year of 2011, the Belle Collaboration reported two charge bottomonium-like state, $Z_b(10610)$ and $Z_b(10650)$, and two year later, a series of charmonium-like states named $Z_c$ which can be considered as the charm analog of $Z_b$ states, have been observed by BESIII and Belle Collaborations. The observations of the $Z_b$ and $Z_c$ states show validity of heavy quark symmetry. Recently, the BESIII Collaboration reported a new charmonium-like structure with strangeness, named $Z_{cs}(3985)$ near the threshold of $D_{s}^{-}D^{*0}/D_{s}^{*-}D^{0}$ in the $K^{+}$ recoil-mass spectrum of the processes  $e^{+}e^{-}\rightarrow K^{+}(D^{-}_{s}D^{*0}+D^{*-}_{s}D^{0})$. And later, the LHCb Collaboration observed a similar states $Z_c(4000)$ but with a large width. The experimental observation stimulate us to investigate the strange hidden-bottom tetraquarks system, which can be considered as the bottom analogy of the newly observed $Z_{cs}(3985)$. In the present work, the strange hidden bottom tetraquarks with $I(J^{P})=\frac{1}{2}(0^{+}), \frac{1}{2}(1^{+})$ and $\frac{1}{2}(2^{+})$ have been systemically investigated in two kinds of quark models, i.e., ChQM and QDCSM. In the estimations, the meson-meson and diquark-antidiquark configurations have been taken into account.

To search for the possible bound states in the strange hidden bottom tetraquark systems, we perform  systematic estimations in the ChQM and QDCSM, where we  evaluate the effect of channels coupling by comparing the results from the single-channel and the coupling-channel estimations. Our single channel estimations indicate that there are no bound states in ChQM, while in QDCSM, the attraction between $s\bar{b}$ and $b\bar{q}$ are strong enough to form bound states, the energy of which is lower than the threshold of the corresponding channels. After considering the coupling channel effects, our estimations indicate the lowest eigenenergy is still above the lowest theoretical threshold, and one can conclude that there is no bound state in both ChQM and QDCSM. However, we notice that the energy of each single channel will rise when the two subclusters are too close in the diquark-antidiquark configuration, so there is a hinderance for the state changing structure to $q\bar{q}-q\bar{q}$ even if the energy of the $qq-\bar{q}\bar{q}$ state is higher than the $q\bar{q}-q\bar{q}$ state, so it is possible to form resonance states in the strange hidden bottom system. Thus, in the present work, we also apply the stabilization method to couple all channels of both configurations in ChQM and QDCSM and  several resonances are predicted. In particular, there are three resonances with $J(J^P)=\frac{1}{2}(0^+)$, three resonances with $J(J^P)=\frac{1}{2}(1^+)$, and one resonance with $J(J^P)=\frac{1}{2}(2^+)$, which may be accessible for the further experiments in LHCb.

\acknowledgments{This work is supported partly by the National Natural Science Foundation of China under
Contract Nos. 12175037, 11775050, 11775118, 11535005, and 11865019, and is also supported by the Fundamental Research Funds for the Central Universities No. 1107022104; and the China Postdoctoral Science Foundation No. 1107020201.
}

\end{document}